\documentclass[twoside]{article}
\usepackage{fleqn,espcrc2}

\usepackage{graphicx}
\usepackage{amsfonts}
\usepackage[mathscr]{eucal} 
\usepackage{caption2} 
\usepackage{psfrag}
\usepackage{fancyhdr,graphicx}
\usepackage{latexsym,amssymb,epsfig}
\usepackage{amsmath,amssymb}

\newcommand{\be}{\begin{equation}}
\newcommand{\ee}{\end{equation}}
\newcommand{\bea}{\begin{eqnarray}}
\newcommand{\eea}{\end{eqnarray}}

\newcommand{\noi}{\noindent}

\newcommand{\fr}{\frac}

\usepackage{axodraw,color}

\title{\Large \bf 
Impact of Duality Violations on Spectral Sum Rule analyses
\thanks{Talk delivered at the QCD 05: 12th High-Energy Physics International Conference 
in Quantum Chromodynamics, Montpellier (France), July 2005.}}

\author{Oscar Cat\`a\address{Grup de F\'isica Te{\`o}rica and IFAE, 
Universitat Aut{\`o}noma de Barcelona, \\
        E-08193 Bellaterra, Barcelona, Spain}}

\begin{document}
\textwidth 16.5cm
\textheight 20.6cm
\evensidemargin -0.2cm
\pagestyle{empty}
\begin{abstract}
Recent sum rule analyses on the $<$VV-AA$>$ two-point correlator have led to significant discrepancies in the values found for the OPE condensates, most dramatically in the dimension eight condensate and to a lesser extent in the dimension six one \cite{condensates}. Precise knowledge of these condensates is of relevance in kaon decays \cite{relationship} and therefore it seems mandatory to assess the actual impact of what is commonly neglected in spectral sum rules, most prominently the issue of duality violations. We will explicitly compute them in a toy model and show that they are {\it{a priori}} non-negligible.

\end{abstract}

\maketitle

\section{INTRODUCTION}
In the absence of a solution to QCD, extracting information about QCD Green functions is a challenging task. Experimentally, only their imaginary parts, {\it{i.e.}}, the spectral functions
\begin{equation}
\rho_{LR}(t)=\fr{1}{\pi}{\mathrm{Im}}\, \Pi_{LR}(t)
\end{equation}
are accessible. A common strategy is to use Cauchy's integral theorem smartly, so that spectral functions can be related to the full Green functions through
\begin{equation}\label{Cauchy}
\int_0^{s_0}\!\!\!\!\!dt\, t^n\,\rho_{LR}(t)=-\fr{1}{2\pi i} \oint_{|q^2|=s_0}\!\!\!\!\!\!\!\!\!\!\!\!dq^2 q^{2n}\,\Pi_{LR}(q^2)
\end{equation}
and replace $\Pi_{LR}(q^2)$ by its Operator Product Expansion, defined as ($Q^2=-q^2$)
\begin{equation}
\Pi_{LR}^{OPE}(Q^2)\,\,\,=_{_{_{\!\!\!\!\!\!\!\!\!\!\!\! Q^2\gg 0}}}\!\!\sum\,\fr{c_{2k}}{Q^{2k}},\,\,\,\, c_{2k}(Q^2)=a_{2k}+b_{2k}\log{\fr{Q^2}{\mu^2}}.
\end{equation}
However, since the OPE is not convergent over the whole $q^2$-plane (most blatantly over the physical axis) the previous replacement generates an error ${\mathcal{D}}^{[n]}(s_0)$ so that Cauchy's theorem now reads 
\begin{equation}\label{CauchyOPE}
\int_0^{s_0}\!\!\!\!\!dt\, t^n\,\rho_{LR}(t)=-\fr{1}{2\pi i} \oint_{|q^2|=s_0}\!\!\!\!\!\!\!\!\!\!\!\!dq^2 q^{2n}\,\Pi_{LR}^{OPE}(q^2)+{\mathcal{D}}^{[n]}(s_0).
\end{equation}
From now on, we will denote each term in the previous equation as 
\begin{equation}\label{makeiteasy}
M_n(s_0)={\mathcal{A}}_n(s_0)+{\mathcal{D}}^{[n]}(s_0).
\end{equation}
The moments $M_n(s_0)$ are extracted from experiment, whereas the OPE integral can be done straightforwardly to yield
\begin{eqnarray}\label{bterms}
(-1)^n{\mathcal{A}}_n(s_0)&=&a_{2n+2}+b_{2n+2}\ \log{\fr{s_0}{\mu^2}}+\nonumber\\
&+&\sum_{k\neq n}^{\infty}\ (-1)^k\ \fr{b_{2k+2}}{|n-k|}s_0^{(n-k)}.
\end{eqnarray} 
The ${\mathcal{D}}^{[n]}(s_0)$ functions account for the difference between $\Pi_{LR}(q^2)$ and $\Pi_{LR}^{OPE}(q^2)$, {\it{i.e.}}, they measure the amount of {\it{duality violations}} \cite{Shifman}. It is worth stressing that, even though present-day analyses do not implement them, as a matter of principle ${\mathcal{D}}^{[n]}(s_0)\neq 0$. In the next section we will attempt to characterize such terms.

\section{WHAT WE EXPECT ON DUALITY VIOLATIONS}
Our discussion will rest upon very mild assumptions, namely that the OPE is an asymptotic expansion and that large-$N_c$ QCD \cite{largeN} is a good approximation to real QCD. In the deep Euclidean regime, a Green function can be approximated by its OPE up to exponentially suppressed terms (the typical uncertainty in asymptotic expansions)
\begin{equation}\label{DV}
\Pi_{LR}(q^2)\approx \Pi_{LR}^{OPE}(q^2) +\mathcal{O}\left(e^{-\fr{2\pi|q^2|}{\lambda^2}}\right),
\end{equation}
$\lambda$ being a typical scale. To be conservative, we consider the OPE to be valid all over the left half complex $q^2$-plane. We can safely extend (\ref{DV}) over the whole $q^2$-plane ($|q^2|\gg0$) but on the physical axis, where singularities are located. This means that over the right half $q^2$-plane, except for the analytically-continued OPE terms there can only appear exponential pieces, which we denote under $\Delta(q^2)$, and  
\begin{equation}\label{mink}
\Pi_{LR}(q^2)\approx \Pi_{LR}^{OPE}(q^2)+\Delta(q^2)+\mathcal{O}\left(e^{-\fr{2\pi|q^2|}{\lambda^2}}\right).
\end{equation}
This extra piece $\Delta(q^2)$ has to depend somehow on the number of colors $N_c$. Recall that in the strict large-$N_c$ limit resonances are infinitely narrow and nothing like an exponential damping is expected over the physical axis. It turns out \cite{Cata} that $\Delta(q^2)$ can be cast in the form
\begin{equation}\label{charac}
\Delta(q^2)\sim {\mathrm{exp}}\left[-2\pi\fr{|q^2|}{\lambda^2}{\mathcal{F}}(\phi,N_c)\right]{\mathcal{H}}(q^2),
\end{equation}
where ${\mathcal{F}}(\phi,N_c)$ is a function such that in the $N_C\longrightarrow \infty$ limit the exponential behaviour no longer holds along the Minkowski axis ($\phi=0,2\pi$). Regarding ${\mathcal{H}}(q^2)$, not much can be said in general. However, bearing in mind that $\Pi_{LR}$ is the difference between vector and axial currents, at least its imaginary part has to be of an oscillatory type.
Comparing (\ref{mink}) and (\ref{CauchyOPE}), one can see that $\Delta(q^2)$ and ${\mathcal{D}}^{[n]}(s_0)$ are intimately related. Actually,
\begin{equation}\label{dualviol}
    \mathcal{D}^{[n]}(s_0)=-\frac{1}{2\pi i}\
    \int_{_{_{_{\!\!\!\!\!\!\!\!\!\!\!\!\!\!\!\!\!\!{\begin{array}{c}
     {|q^2|=s_0} \\ {\rm Re}\ q^2\geq 0
    \end{array}}}}}} \!\!\!\!\!\!\!\!\!\!\!\!\!\!\! dq^2\ q^{2n}\
    \Delta(q^2)
    + \mathcal{O}\left( e^{-\fr{2\pi s_0}{\lambda^2}}\right).
\end{equation}
Use of Cauchy residue theorem bearing in mind the exponential behaviour of (\ref{charac}) allows one to eventually express the previous equation as\footnote{See \cite{Cata} for a detailed derivation.}
\begin{equation}\label{finally}
    \mathcal{D}^{[n]}(s_0)=- \int_{s_0}^{\infty} dt\ t^n\ \frac{1}{\pi}
    \ \mathrm{Im}\ \Delta(t+i\varepsilon)\ +\
    \mathcal{O}\left(e^{-\frac{2 \pi s_0}{\lambda^2}}\right),
\end{equation}
{\it{i.e.}}, only in terms of the imaginary part of (\ref{charac}).
\section{A TOY MODEL}
Lacking knowledge of QCD Green functions, in order to proceed further one has to resort to models, where analytic results are available. Our aim is two-fold: 1) we will be able to test different spectral sum rule techniques and 2) do an explicit calculation of the duality violating pieces.  

We will consider the following {\it{ans\"atze}} for the spectral functions 
\begin{eqnarray}
\rho_V&=&2 F_{\rho}^2
\delta(t-m_{\rho}^2)+ 2 \sum_{n=0}^{\infty} F_V^2\delta(t-M^2_V(n))\nonumber\\
\rho_A&=&2 F_{0}^2 \delta(t)+2 \sum_{n=0}^{\infty}
F_A^2\delta(t-M^2_A(n)), 
\end{eqnarray}
where both axial and vectorial towers display a Regge-like behaviour \cite{regge}
\begin{equation}
M_{V,A}^2(n)=m_{V,A}^2+n\,\Lambda^2\, \quad , \quad
    F^2_{V,A}(n)= F^2.
\end{equation}
For simplicity, we take a single spacing $\Lambda^2$ and decay constant $F^2$. So far, our model displays a spectrum of infinitely narrow vector and axial resonances. We want to provide them with a non-zero width while preserving the analytic properties of $\Pi_{LR}(q^2)$. A simple choice \cite{Blok} is to shift $q^2$ to
\begin{equation}
z=\Lambda^2\left(\fr{-q^2-i\varepsilon}{\Lambda^2}\right)^{\zeta}\, , \qquad \zeta=1-\fr{a}{\pi N_c}.
\end{equation}
With the previous replacement, widths have the expected $N_c$-scaling
\begin{equation}
\Gamma_{V,A}(n)=\frac{a}{N_c} M_{V,A}(n)\sim \fr{\sqrt{n}}{N_c},
\end{equation}
and the full Green function reads
\begin{eqnarray}\label{green}
\Pi_{LR}(q^2)&=&\frac{1}{\zeta}\bigg[
    -\frac{F^2_{0}}{z}+\frac{F_{\rho}^2}{z+m_{\rho}^2}+\nonumber\\
   &\!\!\!\!\!\!\!\!\!\!\!\!\!\!\!\!\!\!\!\!\!\!\!\!\!\!\!\!\!\!\!\!\!\!\!+&\!\!\!\!\!\!\!\!\!\!\!\!\!\!\!\!\!\!\!\!\!\! \frac{F^2}{\Lambda^2}\left\{\psi\left(\frac{z+m_A^2}{\Lambda^2}\right)-
    \psi\left(\frac{z+m_V^2}{\Lambda^2}\right)\right\}\bigg].
\end{eqnarray}
Free parameters are fixed by demanding the short-distance behaviour of QCD to be matched, namely the parton model coefficient for both vector and axial channels and the two Weinberg sum rules. From (\ref{green}) it is straightforward to obtain the OPE coefficients. To leading order in $a/N_c$ they read
\begin{eqnarray}
a_{2k}&\sim& \fr{1}{k}F^2\Lambda^{(2k-2)}\left[B_k\left(\fr{m_V^2}{\Lambda^2}\right)-B_k\left(\fr{m_A^2}{\Lambda^2}\right)\right]\nonumber\\
b_{2k}&=& \fr{ka}{N_c}\,a_{2k}.
\end{eqnarray}
Bernoulli polynomials display (asymptotically) a factorial growth. This signals at an asymptotic OPE for our model, as it is believed to happen in real QCD.
 
The game we want to play now is the following: we will take our toy model as the real world and test the different existing approaches, mainly FESR and pFESR \cite{deRafael} upon it. 
FESR rely on the assumptions that ${\mathcal{A}}_{0,1}=0$ and that there exist duality points $s_0^*$ which lie rather close for neighbouring moments, {\it{i.e.}},
\begin{equation}\label{FESR}
{\mathcal{D}}^{[0,1]}(s_0^*)=0\,\,\Longrightarrow\,\, {\mathcal{D}}^{[2]}(s_0^*)\simeq 0\, , \,\,\,{\mathcal{D}}^{[3]}(s_0^*)\approx 0.
\end{equation}
On the other hand, the pFESR approach relies on a slightly modified version of (\ref{Cauchy}), namely
\begin{equation}\label{Cauchypinch}
\int_0^{s_0}\!\!\!\!\!dt\, w(t)\,\rho_{LR}(t)=-\fr{1}{2\pi i} \oint_{|q^2|=s_0}\!\!\!\!\!\!\!\!\!\!\!\!dq^2 w(q^2)\,\Pi_{LR}(q^2),
\end{equation}
where $w(t)$ are polynomials which, because they vanish for $t=s_0$, are expected to minimize duality violations. In pFESR, one has a combination of moments that has to match the corresponding combination of OPE coefficients. Usually, such OPE coefficients are fixed by fitting both curves over a window in $s_0$.

However, once applied to our model, both FESR and pFESR yield predictions with huge errors when compared to the true OPE coefficients. We definitely have to do better and assess the impact of the neglected terms. If we take the imaginary part of (\ref{mink}),
\begin{equation}
{\mathrm{Im}}\,\Pi_{LR}(t)={\mathrm{Im}}\Pi_{LR}^{OPE}(t)+{\mathrm{Im}}\Delta(t),
\end{equation}
and use our model to explicitly compute the different pieces we end up with \cite{Cata}
\begin{eqnarray}\label{opeim}
{\mathrm{Im}}\Pi_{LR}^{OPE}(t)&=&\fr{3a}{N_c}\fr{b_6}{t^3}+{\mathcal{O}}(t^{-4})\\  
{\mathrm{Im}}\Delta(t)&=& \kappa\, e^{-\gamma t}\, \sin{(\alpha+\beta t)}\label{dual},
\end{eqnarray}
where
\[
\kappa=\fr{4\pi F^2}{\Lambda^2}\sin{\left(\pi\fr{m_V^2-m_A^2}{\Lambda^2}\right)}\, , \quad
\gamma=\fr{2\pi a}{N_c\Lambda^2}
\]
\begin{equation}
\alpha=-\fr{\pi}{\Lambda^2}(m_V^2+m_A^2)\, , \quad
\beta=\fr{2\pi}{\Lambda^2}.
\end{equation}
Eq. (\ref{dual}) is nothing but a particular example of what we already stated in (\ref{charac}), where ${\mathrm{Im}}\,\,{\mathcal{H}}(q^2)$ is, to a very good approximation, of a sinusoidal type. Eq. (\ref{opeim}) is the analytical continuation of the OPE into the Minkowski region. Therefore, the absence of $t^{-1}$ and $t^{-2}$ powers is a direct consequence of the Weinberg sum rules we enforced earlier on. Those power corrections are closely related to the second line of (\ref{bterms}) and have a (in principle) non-negligible impact, presumably already at dimension eight, and surely for higher order condensates, where they have positive powers of $s_0$ out front. Both duality violations and power corrections (\ref{bterms}) are generic and nonetheless missing in all existing analyses \cite{condensates}. Any strategy for improvement should take them both into account.

One possibility \cite{Cata} is to take the first two moments in (\ref{makeiteasy}), whose explicit form is
\begin{eqnarray}
M_0(s_0)&=&\left[-\fr{b_6}{2s_0^2}+\fr{b_8}{3s_0^3}+\cdots\right]-\int_{s_0}^{\infty}dt{\mathrm{Im}}\Delta(t)\nonumber\\
M_1(s_0)&=&\left[-\fr{b_6}{s_0}+\fr{b_8}{2s_0^2}+\cdots\right]-\int_{s_0}^{\infty}dt\,t{\mathrm{Im}}\Delta(t),\nonumber\\
&&
\end{eqnarray}
with ${\mathrm{Im}}\,\Delta(q^2)$ as given by (\ref{dual})\footnote{The use of a sinusoidal function instead of the (unknown) ${\mathrm{Im}}\,\,{\mathcal{H}}(q^2)$ is a justified approximation as long as one does the fit over not too large a window.}
and determine the parameters  $\kappa$, $\gamma$, $\alpha$ and $\beta$ through a fit. In the above equations, terms coming from the OPE are strongly suppressed: first, Weinberg sum rules ensure that $c_2$ and $c_4$ vanish, and it seems reasonable to neglect (at first) the remaining $b_{2k}$-terms due to the suppressing powers of $s_0$. Then a prediction for ${\mathcal{A}}_{2,3}$ is possible through
\[
M_2(s_0)+\int_{s_0}^{\infty}dt\,t^2{\mathrm{Im}}\Delta(t)=\left[c_6(s_0)+\fr{b_8}{s_0}+\cdots\right]
\]
\begin{eqnarray}
M_3(s_0)+\int_{s_0}^{\infty}dt\,t^3{\mathrm{Im}}\Delta(t)\!\!\!\!\!&=&\!\!\!\!\!\left[-c_8(s_0)+b_6s_0-\fr{b_{10}}{s_0}+\cdots\right]\nonumber\\
&&
\end{eqnarray}
for a given $s_0$. Notice that this strategy shares some similarities with FESR, but we need no longer assume the same duality points for different moments: we can determine the OPE coefficients, {\it{e.g.}}, through a fit in a window of $s_0$. In order to make a comparison, we list on the table below \cite{Cata} the results found {\it{in our model}}, where one can see that the {\it{duality violation approach}} (DVA) we propose outdoes conventional FESR and pFESR.
\begin{center}
\begin{tabular}{|c|c|c|}
\hline
 & ${\mathcal{A}}_6\,(GeV^6)$ & ${\mathcal{A}}_8\,(GeV^8)$\\
\hline \hline
FESR &$-2.0 \cdot 10^{-3}$  & $-1.6 \cdot 10^{-3}$ \\
\hline
pFESR & $-3.8 \cdot 10^{-3}$ & $+6.5 \cdot 10^{-3}$ \\
\hline
DVA & $-2.51\cdot 10^{-3}$  & $+3.29\cdot 10^{-3}$\\
\hline
OPE & ${\mathbf{-2.81\cdot 10^{-3}}}$  & ${\mathbf{+3.44\cdot 10^{-3}}}$ \\
\hline
\end{tabular}
\end{center}

\section{CONCLUSIONS}
Based on rather mild assumptions, namely that the OPE is an asymptotic expansion and that the $1/N_c$ expansion is a good approximation to QCD, we would like to propose a general expression for the duality violating pieces in $\Pi_{LR}(q^2)$. With the help of a toy model inspired by large-$N_c$ we have been able to compute duality violations explicitly. Then, using the model as if it were the real world, we have tested FESR and pFESR upon it. Both approaches yielded predictions for the condensates with large errors, showing that duality violations might have a bigger impact than commonly expected. We propose an analysis that does include duality violations and in which errors reduce down to typically 10-15$\%$, which is a clear improvement over FESR and pFESR. Interestingly, the method can also be applied to QCD, since it relies on very generic assumptions. Finally, we would like to emphasize that duality violations are not the only source of uncertainty in present-day analyses: one should not forget about potentially harmful $s_0$ power corrections (see second line of (\ref{bterms})), whose impact might be non-negligible already at the dimension eight condensate.    

\vspace*{7mm} 

{\large{\bf Acknowledgments}}

\vspace*{3mm}

\noi I want to thank Maarten Golterman and Santi Peris for a very pleasant collaboration. I also want to thank the people at the Benasque Center for Science for their hospitality. And last but not least, the organizers of the QCD 05 Conference for their kind invitation. This work was supported in part by CICYT-FEDER-FPA2002-00748, 2001-SGR00188 and by TMR, EC-Contract No. HPRN-CT-2002-00311 (EURIDICE).

\end{document}